\documentclass[11pt,a4paper]{article}
 \usepackage{amsmath,amssymb}
  \usepackage[dvips,final]{graphicx}
   \graphicspath{{./bakulev-figs/}}

\begin{document}

\title{\textbf{Two-loop Resummation in Fractional Analytic Perturbation Theory}}
\author{Alexander~P.~Bakulev$^a$\footnote{\textbf{E-mail}: bakulev@theor.jinr.ru}
\\
$^a$ \small{\em Bogoliubov Laboratory of Theoretical Physics, JINR} \\
\small{\em Jolio-Curie 6, Dubna, 141980 Russia}
}
\date{\today}
\maketitle

\begin{abstract}
 This talk describes the resummation approach in (Fractional) Analytic Perturbation Theory
 (FAPT) in QCD.
 First, we make a short historical review of the (F)APT approach
 and then shortly describe the global scheme of FAPT
 which allows one to take into account heavy-quark thresholds.
 After that we show how it is possible to resum a non-power series
 in (F)APT both in the one- and two-loop approximations.
 As an application we suggest our analysis
 of the Higgs boson decay ${H^0\to b\overline{b}}$,
 important for the LHC program,
 and of the vector-current Adler function.
\end{abstract}

\section{Analytic Perturbation Theory in QCD}
 \label{sec:APT}

First, I'd like to remind the history of Analytic Perturbation Theory (APT).
Strictly speaking,
APT was initiated by the paper of N.~N.~Bogolyuov et al.
in 1959~\cite{BLS60},
where ghost-free effective coupling for QED was constructed
using the dispersion-relations method.
Then, in 1982,
Radyushkin \cite{Rad82}, and Krasnikov and Pivovarov \cite{KP82},
using the same dispersion-relations technique,
suggested regular (for $s\geq \Lambda^2$) QCD running coupling
in the Minkowski region,
$\alpha_s(s)\equiv a_s(s)/\beta_f$,
namely, $a_s[L]=\pi^{-1}\arctan(\pi/L)$
with
$L=\ln(s/\Lambda^2)$ and
$\beta_f=b_0(N_f)/(4\pi)=(11-2N_f/3)/(4\pi)$.\footnote{%
   We use the notations $f(Q^2)$ and $f[L]$ in order to specify the arguments we mean ---
   squared momentum $Q^2$ or its logarithm $L=\ln(s/\Lambda^2)$ or $\ln(Q^2/\Lambda^2)$,
   that is $f[L]=f(\Lambda^2\cdot e^L)$ and $\Lambda^2$ is usually referred
   to the $N_f=3$ region.}
(This normalized coupling $a_s[L]$ in standard QCD PT
 is just $1/L$ in the one-loop approximation.)

Thirteen years later,
Dokshitzer et al.
constructed an IR finite ${\alpha_s^\textbf{eff}(Q^2)}$ in the Euclidean domain
using a renormalon-based approach with dispersion relations~\cite{DMW96}.
After that, Jones and Solovtsov~\cite{JS95-349,JS95-357}
discovered the coupling
which appears to be finite for all $s$
and coincides with the Radyushkin's one for $s\geq\Lambda^2$,
namely
\begin{subequations}
 \label{eq:A.U}
 \begin{eqnarray}
 \label{eq:U_1}
 {\mathfrak A}_1[L]
  \!\!&\!\!=\!\!&\!\! \int_s^{\infty}\!\frac{\rho_1(\sigma)}{\sigma}\,d\sigma\
   =\ \frac{1}{\pi}\,\arccos\frac{L}{\sqrt{\pi^2+L^2}}\,.
 \end{eqnarray}
Just at the same time,
Ball, Beneke and Braun
reproduced in the pQCD-based renormalon approach
$a_s[L]=\pi^{-1}\arctan(\pi/L)$
in the low-energy Minkowski region~\cite{BB95,BBB95},
and
Shirkov and Solovtsov \cite{SS},
using the same dispersion-relations approach of~\cite{BLS60},
discovered the ghost-free coupling,
\begin{eqnarray}
  \label{eq:A_1}
 \mathcal A_1[L]
  \!\!&\!\!=\!\!&\!\! \int_0^{\infty}\!\frac{\rho_1(\sigma)}{\sigma+Q^2}\,d\sigma\
   =\ \frac{1}{L} - \frac{1}{e^L-1}\,,
\end{eqnarray}
\end{subequations}
in the Euclidean region with $L=\ln(Q^2/\Lambda^2)$.
But Shirkov--Solovtsov approach, named APT,
appears to be more powerful:
in the Euclidean domain,
$\displaystyle-q^2=Q^2$, $\displaystyle L=\ln Q^2/\Lambda^2$,
it generates the set of images for the effective coupling
and its $n$-th powers,
$\displaystyle\left\{{\mathcal A}_n[L]\right\}_{n\in\mathbb{N}}$,
whereas in the Minkowski domain,
$\displaystyle q^2=s$, $\displaystyle L=\ln s/\Lambda^2$,
it generates another set,
$\displaystyle\left\{{\mathfrak A}_n[L]\right\}_{n\in\mathbb{N}}$
(see also in~\cite{Sim01}).
APT is based on the Renormalization Group (RG)
and causality
that guarantees standard perturbative UV asymptotics
and spectral properties.
The power series $\sum_{m}d_m a_s^m[L]$
transforms into a non-power series
$\sum_{m}d_m {\mathfrak A}_{m}[L]$ or $\sum_{m}d_m {\mathcal A}_{m}[L]$ in APT.
The global version of APT,
which takes into account the heavy-quark thresholds,
has been elaborated in~\cite{DVS-Global}.

By the analytization in APT for an observable $f(Q^2)$
we mean the ``K\"allen--Lehmann'' representation
\begin{subequations}
 \label{eq:APT.main}
\begin{eqnarray}
 \label{eq:An.SD}
  \left[f(Q^2)\right]_\text{an}
   = \int_0^{\infty}\!
      \frac{\rho_f(\sigma)}
         {\sigma+Q^2-i\epsilon}\,
       d\sigma
\end{eqnarray}
with $\displaystyle\rho_f(\sigma)=\frac{1}{\pi}\,\textbf{Im}\,\big[f(-\sigma)\big]$,
which transforms to different prescriptions for couplings
in the Euclidean and Minkowski domains
\begin{eqnarray}
 \label{eq:A_n}
  \mathcal A_n[L]
   \!&\!=\!&\!
    \textbf{A}_\textbf{E}\left[\alpha_s^n\right]
     = \int_0^{\infty}\!\frac{\rho_n(\sigma)}{\sigma+Q^2}\,d\sigma\,;\\
 \label{eq:U_n}
  \mathfrak A_n[L]
   \!&\!=\!&\!
    \textbf{A}_\textbf{M}\left[\alpha_s^n\right]
     = \int_s^{\infty}\!\frac{\rho_n(\sigma)}{\sigma}\,d\sigma\,.
\end{eqnarray}
\end{subequations}

In the one-loop approximation,
we have then
$\rho_1(\sigma)=1/\sqrt{L_\sigma^2+\pi^2}$ and Eqs.\,(\ref{eq:A.U}),
whereas analytic images of the higher powers ($n\geq2, n\in\mathbb{N}$)
are~\cite{Rad82,BRS00}
\begin{eqnarray}
 \label{eq:recurr.1L}
 {\mathcal A_n[L] \choose \mathfrak A_n[L]}
   = \frac{1}{(n-1)!}\left( -\frac{d}{d L}\right)^{n-1}
      {\mathcal A_{1}[L] \choose \mathfrak A_{1}[L]}\,.
\end{eqnarray}

At first glance, the APT is a complete theory
supplying tools to produce
an analytic answer for any perturbative series in QCD.
But in 2001 Karanikas and Stefanis
reminded the need for more tools to produce
analytic expressions for hadronic observables,
calculated perturbatively~\cite{KS01}.
Indeed, in the standard QCD PT one has also:
\begin{enumerate}
  \item[(i)] the factorization procedure in QCD
    that gives rise to the appearance of logarithmic factors of the type:
     $a_s^\nu[L]\,L$;
  \item[(ii)] the RG evolution
     that generates evolution factors of the type
     $B(Q^2)=\left[Z(Q^2)/Z(\mu^2)\right]$ $B(\mu^2)$,
     which reduce in the one-loop approximation to
     $Z(Q^2) \sim a_s^\nu[L]$ with $\nu=\gamma_{0}/(2b_0)$
     being a fractional number
     (here the subscript $_0$ means that the corresponding quantity
      is calculated in the one-loop approximation).
\end{enumerate}
They suggested the principle of analytization ``as a whole''
in the $Q^2$ plane for hadronic amplitudes,
obtained in QCD PT,
thus generalizing the analytic approach
suggested in~\cite{SSK9900}.\footnote{%
More precisely, they proposed the analytization recipe
for terms like
$\int_{0}^{1}\!dx\!\int_{0}^{1}\!dy\,
  \alpha_\text{s}\left(Q^{2}xy\right) f(x)f(y)$,
which can be treated as an effective account
for the logarithmic terms
in the next-to-leading-order
approximation of perturbative QCD.}
A review of both approaches
can be found in~\cite{Ste02},
whereas an extensive discussion of the APT application
to the pion form factor analysis
is contained in~\cite{BPSS04}.

All that means
that in order to generalize APT
in the ``analytization as a whole'' direction
one needs to construct analytic images
of new functions:
$\displaystyle a_s^\nu,~a_s^\nu\,L^m, \ldots$\,.
This task has been performed in the framework of the so-called FAPT,
suggested in~\cite{BMS-APT,BKS05}.
Now we briefly describe this approach.
In the one-loop approximation,
using the recursive relation (\ref{eq:recurr.1L}),
we can obtain explicit expressions for
${\mathcal A}_{\nu}[L]$
and ${\mathfrak A}_{\nu}[L]$:
\begin{subequations}
\label{eq:recur.1L}
\begin{eqnarray}
 {\mathcal A}_{\nu}[L]
 \!\!&\!\!=\!\!&\!\!
   \frac{1}{L^\nu}
  - \frac{F(e^{-L},1-\nu)}{\Gamma(\nu)}\,;
 ~
 \\
 {\mathfrak A}_{\nu}[L]
 \!\!&\!\!=\!\!&\!\!
   \frac{\text{sin}\left[(\nu -1)\arccos\left(\frac{L}{\sqrt{\pi^2+L^2}}\right)\right]}
         {\pi(\nu -1) \left(\pi^2+L^2\right)^{(\nu-1)/2}}\,.~
\end{eqnarray}
\end{subequations}
Here $F(z,\nu)$ is the reduced Lerch transcendental function,
which is an analytic function in $\nu$.
This function has very interesting properties,
which were discussed extensively
in the previous papers~\cite{BMS-APT,BKS05,AB08,Ste09}.

 The construction of FAPT with a fixed number of quark flavors, $N_f$,
is a two-step procedure:
we start with the perturbative result $\left[a_s(Q^2)\right]^{\nu}$,
generate the spectral density
using
$\displaystyle\rho_{\nu}(\sigma)=\frac{1}{\pi}\,\textbf{Im}\,\big[a_s(-\sigma)\big]^{\nu}$,
and then obtain analytic couplings
${\mathcal A}_{\nu}[L]$ and ${\mathfrak A}_{\nu}[L]$
via the relations like Eqs.\ (\ref{eq:APT.main}).
Here $N_f$ is fixed and factorized out.
We can proceed in the same manner with $N_f$-dependent quantities:
$\left[\alpha_s^{}(Q^2;N_f)\right]^{\nu}$
$\Rightarrow$
$\overline{\rho}_{\nu}(\sigma;N_f)=\overline{\rho}_{\nu}[L_\sigma;N_f]
 \equiv\rho_{\nu}(\sigma)/\beta_f^{\nu}$
$\Rightarrow$
$\overline{\mathcal A}_{\nu}^{}[L;N_f]$ and $\overline{\mathfrak A}_{\nu}^{}[L;N_f]$ ---
here $N_f$ is fixed, but not factorized out.
The global version of FAPT~\cite{AB08},
which takes into account heavy-quark thresholds,
is constructed along the same lines
but starting from the global perturbative coupling
$\left[\alpha_s^{\,\text{\tiny glob}}(Q^2)\right]^{\nu}$,
being a continuous function of $Q^2$
due to adopting different values of QCD scales $\Lambda_f$,
which correspond to different values of $N_f$.
We illustrate here the case of only one heavy-quark threshold
at $s=m_4^2$,
corresponding to the transition $N_f=3\to N_f=4$.
Then we obtain the discontinuous spectral density
\begin{eqnarray}
 \rho_n^\text{\tiny glob}(\sigma)
   = \theta\left(L_\sigma<L_{4}\right)\,
       \overline{\rho}_n\left[L_\sigma;3\right]
   + \theta\left(L_{4}\leq L_\sigma\right)\,
       \overline{\rho}_n\left[L_\sigma+\lambda_4;4\right]
 \label{eq:global_PT_Rho_4}
\end{eqnarray}
with $L_{\sigma}\equiv\ln\left(\sigma/\Lambda_3^2\right)$,
$L_{f}\equiv\ln\left(m_f^2/\Lambda_3^2\right)$
and
$\lambda_f\equiv\ln\left(\Lambda_3^2/\Lambda_f^2\right)$ for $f=4$,
which is expressed in terms of fixed-flavor spectral densities
with 3 and 4 flavors,
$\overline{\rho}_n[L;3]$ and $\overline{\rho}_n[L+\lambda_4;4]$.
However, it generates the continuous Minkowski coupling
\begin{subequations}
\begin{eqnarray}
 {\mathfrak A}_{\nu}^{\text{\tiny glob}}[L]
  = \theta\left(L\!<\!L_4\right)
     \Bigl(\overline{{\mathfrak A}}_{\nu}^{}[L;3]
          + \Delta_{43}\overline{{\mathfrak A}}_{\nu}^{}
     \Bigr)
  + \theta\left(L_4\!\leq\!L\right)\,
     \overline{{\mathfrak A}}_{\nu}^{}[L+\lambda_4;4]
 \label{eq:An.U_nu.Glo.Expl}
\end{eqnarray}
with $\Delta_{43}\overline{{\mathfrak A}}_{\nu}^{}=
            \overline{{\mathfrak A}}_{\nu}^{}[L_4+\lambda_4;4]
          - \overline{{\mathfrak A}}_{\nu}^{}[L_4;3]$
and
the analytic Euclidean coupling ${\cal A}_{\nu}^{\text{\tiny glob}}[L]$
\begin{eqnarray}
 {\cal A}_{\nu}^{\text{\tiny glob}}[L]
  = \overline{{\cal A}}_{\nu}^{}[L+\lambda_4;4]
  + \int\limits_{-\infty}^{L_4}\!
       \frac{\overline{\rho}_{\nu}^{}[L_\sigma;3]
            -\overline{\rho}_{\nu}^{}[L_\sigma+\lambda_{4};4]}
            {1+e^{L-L_\sigma}}\,
         dL_\sigma\,.
  \label{eq:Delta_f.A_nu}
\end{eqnarray}
\end{subequations}

The same strategy one needs to use
when working in a higher-loop approximation ---
then, analytic formulas even for the fixed-$N_f$ case
are unavailable
(the spectral density is expressed
 in terms of the imaginary part of the Lambert function)
and one is forced to use integral representations of the type (\ref{eq:APT.main})
to calculate analytic couplings and other quantities of interest.
One more complication here is the more complex evolution law:
for example, in the two-loop approximation the evolution factor
$B(Q^2)=\left[Z(Q^2)/Z(\mu^2)\right]$ has
$Z(Q^2) \sim a_s^{\nu_0}[L](1+c_1a_s[L])^{\nu_1}$,
with exponents
$\nu_{0}$ and $\nu_{1}$
being known numerical coefficients ---
for more details see in~\cite{AB08}.

\section{Resummation in (F)APT}
\label{sec:Resum.FAPT}

We consider now the perturbative expansion
of a typical physical quantity,
like the Adler function and the ratio $R$,
in the one-loop APT.
Due to the limited space of our presentation,
we provide all formulas only
for quantities in the Minkowski region:
\begin{subequations}
\begin{eqnarray}
 \label{eq:APT.Series.Gen}
  \mathcal R[L]
   = \sum_{n=1}^{\infty}
      d_n\,\mathfrak A_{n}[L]\,.
\end{eqnarray}
We suggest that there exist a generating function $P(t)$
for the coefficients $\tilde{d}_n=d_n/d_1$:
\begin{equation}
 \tilde{d}_n
  =\int_{0}^\infty\!\!P(t)\,t^{n-1}dt
   ~~~\text{with}~~~
   \int_{0}^\infty\!\!P(t)\,d t = 1\,.
 \label{eq:generator}
\end{equation}
To shorten our formulae, we use for the integral
$\displaystyle\int\limits_{0}^{\infty}\!\!f(t)P(t)dt$
the following notation:
$\langle\langle{f(t)}\rangle\rangle_{P(t)}$.
Then, the coefficients $d_n = d_1\,\langle\langle{t^{n-1}}\rangle\rangle_{P(t)}$
and the series (\ref{eq:APT.Series.Gen})
can be represented as follows (we put for shortness $d_1=1$):
\begin{eqnarray}
 \label{eq:APT.Series}
  \mathcal R[L]
   = \sum_{n=1}^{\infty}
      \langle\langle{t^{n-1}}\rangle\rangle_{P(t)}\,
       \mathfrak A_{n}[L]\,.
\end{eqnarray}
\end{subequations}

 \subsection{One-loop case}
  \label{sec:Resum.FAPT.1L}

Due to the recurrence relations (\ref{eq:recur.1L}),
as has been shown in~\cite{MS04},
we have the exact result for the sum in (\ref{eq:APT.Series}),
viz.,
\begin{eqnarray}
 \label{eq:APT.Sum.DR[L]}
  \mathcal R[L]
   = \langle\langle{\mathfrak A_1[L-t]}\rangle\rangle_{P(t)}\,.
\end{eqnarray}
The integral over the variable $t$ has a rigorous meaning,
ensured by the finiteness of the coupling  $\mathfrak A_1[t] \leq 1$
and the fast fall-off of the generating function $P(t)$.
This result resembles the result of resummation
of the large
$(\beta_0^n\alpha_s)^{n-1}$ terms due to Neubert~\cite{Neu95prd},
but here we have no problem with renormalon poles,
i.\,e.,
in our case the integral
$\langle\langle{\mathfrak A_1[L-t]}\rangle\rangle_{P(t)}$
is nicely convergent.
Note
that the same type of approach in the one-loop APT
has been independently
invented two years later\footnote{%
Note that the paper of Mikhailov~\cite{MS04}
was put into arXiv in November, 2004,
and after a long discussions with referees
published only in 2007,
whereas the paper of Cvetic and Valenzuela~\cite{CV06}
was put in arXiv in August, 2006,
and published the same year.
It is interesting to remind here
the title of Mikhailov's paper:
``\textit{Generalization of BLM procedure and its scales
in any order of pQCD: A practical approach}'' ---
we see that it was devoted to completely different subject
and Eq.\,(\ref{eq:APT.Sum.DR[L]}), numerated in this paper as Eq.\,(2.7),
was obtained in passing.
} by Cvetic and Valenzuela~\cite{CV06}
within the framework of the so-called ``skeleton expansion''.

We have first accomplished the generalization
of result (\ref{eq:APT.Sum.DR[L]})
to the case of global APT~\cite{DVS-Global},
taking into account heavy-quark thresholds~\cite{AB08,BM08}.
Then, one starts with the series
of type (\ref{eq:APT.Series}),
where $\mathfrak A_{n}[L]$
are substituted by their global analogs
$\mathfrak A_{n}^\text{\tiny glob}[L]$
(note that due to the different normalizations
 of the global couplings,
 $\mathfrak A_{n}^\text{\tiny glob}[L]\simeq\mathfrak A_{n}[L]/\beta_f$,
 the coefficients $d_n$ should also be changed).
Then, we have
\begin{eqnarray}
 \mathcal R^\text{\tiny glob}[L]
  = \theta(L\!<\!L_4)
     \langle\langle{
      \Delta_{4}\overline{\mathfrak A}_{1}[t]
     + \overline{\mathfrak A}_{1}\!\Big[L\!-\!\frac{t}{\beta_3};3\Big]
     }\rangle\rangle_{P(t)}
  + \theta(L\!\geq\!L_4)
     \langle\langle{
      \overline{\mathfrak A}_{1}\!\Big[L\!+\!\lambda_4-\!\frac{t}{\beta_4};4\Big]
     }\rangle\rangle_{P(t)}\,,~
 \label{eq:sum.R.Glo.4}
\end{eqnarray}
where $\Delta_4\overline{\mathfrak A}_\nu[t]\equiv
  \overline{\mathfrak A}_\nu\!\Big[L_4+\lambda_{4}-t/\beta_4;4\Big]
 -\overline{\mathfrak A}_\nu\!\Big[L_3-t/\beta_3;3\Big]$.

The second generalization has been obtained for the case
of global FAPT.
The starting point in this case
is
the generalized recurrence relation
\begin{eqnarray}
 \label{eq:recurr.FAPT.1L}
  -\frac{1}{n+\nu}\, \frac{d}{dL}\, \mathcal F_{n+\nu}[L]
  = \mathcal F_{n+1+\nu}[L]\,,
\end{eqnarray}
where
${\mathcal F}[L]$ denotes one of the analytic quantities
$\mathcal A[L]$ or $\mathfrak A[L]$.
The result of summation of the series
$\sum_{n=0}^{\infty} \langle\langle{t^{n-1}}\rangle\rangle_{P(t)}\,
 \mathfrak A_{n+\nu}^\text{\tiny glob}[L]$
is the complete analog of Eq.\ (\ref{eq:sum.R.Glo.4})
with the substitutions
\begin{eqnarray}
 \label{eq:P_nu(t)}
  P(t)\Rightarrow P_{\nu}(t) =
   \int_0^{1}\!P\left(\frac{t}{1-x}\right)
    \frac{\nu\,x^{\nu-1}dx}
         {1-x}\,,
\end{eqnarray}
where
$d_0\Rightarrow d_0\,\overline{\mathfrak A}_{\nu}[L]$,
$\overline{\mathfrak A}_{1}[L-t]\Rightarrow
 \overline{\mathfrak A}_{1+\nu}[L-t]$,
and
$\Delta_4\overline{\mathfrak A}_{1}[t]\Rightarrow
 \Delta_4\overline{\mathfrak A}_{1+\nu}[t]$.
All needed formulas have also been obtained
in parallel for the Euclidean case,
see in~\cite{AB08}.

 \subsection{Two-loop case}
  \label{sec:Resum.FAPT.2L}

  In the case of the two-loop running of $\alpha_s^{(2)}(Q^2)=4\,\pi\,a_{(2)}(Q^2)/b_0$,
i.\,e.,
when the $\beta$-function has the following form
(with $c_1\equiv b_1/b_0^2$)
\begin{eqnarray}
 \label{eq:alpha.2L}
  \beta(a_{(2)})
   = -a_{(2)}^2\,
       \left[1 + c_1\,a_{(2)}\right]\,,
\end{eqnarray}
we have two types of complication.
First, the recurrence relations (\ref{eq:recurr.1L}) and (\ref{eq:recurr.FAPT.1L})
transform into
\begin{eqnarray}
 \label{eq:recurr.FAPT.2L}
  -\frac{1}{n+\nu}\, \frac{d}{dL}\, \mathcal F_{n+\nu}[L]
  = \mathcal F_{n+1+\nu}[L]
  + c_1\,\mathcal F_{n+2+\nu}[L]\,,
\end{eqnarray}
where ${\mathcal F}[L]$
denotes now one of the analytic quantities
$\mathcal A^{(2)}[L]$, $\mathfrak A^{(2)}[L]$, or $\rho^{(2)}[L]$.
Second, the evolution factors appear to be more complicated as well,
namely,
\begin{eqnarray}
 \label{eq:evol.2L}
  \mathcal Z_{\nu_0;\nu_1}[L]
   = a_{(2)}^{\nu_0}[L]
     \left(1+c_1\,a_{(2)}\right)^{\nu_1}[L]\,,
\end{eqnarray}
with exponents
$\nu_{0}$ and $\nu_{1}$
being numerical combinations of
anomalous-dimension coefficients $\gamma_0$, $\gamma_1$,
and $\beta$-function coefficients $\beta_0$, $\beta_1$.

For these reasons,
we need to consider the resummation of the following series:
\begin{subequations}
\begin{eqnarray}
 \label{eq:FAPT.series.2L}
  \mathcal S_{\nu}[L]
   &=& \sum_{n=1}^{\infty}
        \langle\langle{t^{n-1}}\rangle\rangle_{P(t)}\,
         \mathcal F_{n+\nu}[L]\,;\\
 \label{eq:FAPT.series.2L.Evo}
  \mathcal S_{\nu_0,\nu_1}[L]
   &=& \sum_{n=1}^{\infty}
        \langle\langle{t^{n-1}}\rangle\rangle_{P(t)}\,
         \mathcal F_{n+\nu_0,\nu_1}[L]\,.
\end{eqnarray}
In the last line,
one has
$\mathcal F_{n+\nu_0,\nu_1}[L]=\mathcal B_{n+\nu_0,\nu_1}^{(2)}[L]$
or
$\mathfrak B_{n+\nu_0,\nu_1}^{(2)}[L]$
(or---in the case of global FAPT---the spectral density $\rho_{n+\nu_0,\nu_1}^{(2)}[L]$
 for these functions)
with
\begin{eqnarray}
 \label{eq:B.nu.nu1}
  {\mathcal B_{\nu_0;\nu_1}[L]\choose\mathfrak B_{\nu_0;\nu_1}[L]}
   = {\textbf{A}_\textbf{E}\choose\textbf{A}_\textbf{M}}
      \left[a_{(2)}^{\nu_0}
             \left(1+c_1\,a_{(2)}\right)^{\nu_1}
      \right][L]
\end{eqnarray}
\end{subequations}
being the analytic images of the two-loop evolution factors
in the Euclidean and Minkowski regions,
correspondingly.
These problems have been successfully resolved
in our recent paper~\cite{BMS10}
and here we show only the results.

First, we need to define the two-loop evolution ``logarithmic time'' $\tau(t)$:
\begin{eqnarray}
 \label{eq:2L.time}
  \tau(t)
   \equiv t - c_1\,\ln\left[1+\frac{t}{c_1}\right]\,.
\end{eqnarray}
Then the two-loop FAPT resummation procedure
produces the following answers:\\
For the series (\ref{eq:FAPT.series.2L})
the resummed expression is
\begin{subequations}
\begin{eqnarray}
  \mathcal S_\nu[L]
   \!&\!=\!&\!
    \left\langle\!\!\!\left\langle
     \mathcal F_{1+\nu}[L]
      -\frac{t^2}{c_1+t}\,
       \int_{0}^{1}\!z^\nu dz\,
        \dot{\mathcal F}_{1+\nu}[L+\tau(t\,z)-\tau(t)]
    \right\rangle\!\!\!\right\rangle_{\!\!\!P(t)}
\nonumber
\\ \!&\!+\!&\!
    \left\langle\!\!\!\left\langle
     \frac{c_1\,t}{c_1+t}\,
      \left\{\mathcal F_{2+\nu}[L]
            -\int_{0}^{1}\!dz\,
              \frac{t^2\,z^{\nu+1}}{c_1+t\,z}\,
               \dot{\mathcal F}_{2+\nu}[L+\tau(t\,z)-\tau(t)]
      \right\}\!
      \right\rangle\!\!\!\right\rangle_{\!\!\!P(t)},~~~
 \label{eq:Resum.2L.FAPT}
\end{eqnarray}
whereas
for the series (\ref{eq:FAPT.series.2L.Evo}) ---
\begin{eqnarray}
 \mathcal S_{\nu_0,\nu_1}[L]
  \!&\!=\!&\!
   \left\langle\!\!\!\left\langle
    \mathcal B_{1+\nu_0;\nu_1}[L]
    + \delta_{0,{\nu_0+\nu_1}}\,
       t\,\left[\frac{c_1}{c_1+t}\right]^{1-\nu_1}
        \mathcal B_{2+\nu_0;\nu_1}[L-\tau(t)]
   \right\rangle\!\!\!\right\rangle_{\!\!\!P(t)}\nonumber\\
  \!&\!-\!&\!
   \left\langle\!\!\!\left\langle
    \frac{t^2}{(c_1+t)^{1-\nu_1}}\,
     \int\limits_{0}^{1}\!dz\,
      \frac{z^{\nu_0+\nu_1}}{(c_1+t\,z)^{\nu_1}}
       \frac{d\mathcal B_{1+\nu_0;\nu_1}[L+\tau(t\,z)-\tau(t)]}{dL}
   \right\rangle\!\!\!\right\rangle_{\!\!\!P(t)}~~~
\nonumber\\
  \!&\!+\!&\!
   \left\langle\!\!\!\left\langle
    \frac{c_1\,t}{(c_1+t)^{1-\nu_1}}\,
     \int\limits_{0}^{1}\!dz\,
      \frac{(\nu_0+\nu_1)\,z^{\nu_0+\nu_1-1}}{(c_1+t\,z)^{\nu_1}}\,
       \mathcal B_{2+\nu_0;\nu_1}[L+\tau(t\,z)-\tau(t)]
  \right\rangle\!\!\!\right\rangle_{\!\!\!P(t)},~~~
 \label{eq:Resum.2L.Evo.imp}
\end{eqnarray}
with  $\delta_{0,{\nu_0+\nu_1}}$ being a Kronecker delta symbol.
\end{subequations}

\section{Applications to Higgs boson decay}
\label{sec:Appl.Higgs}
Here we analyze the Higgs boson decay to a $\overline{b}b$ pair.
For its width we have
\begin{eqnarray}
 \label{eq:Higgs.decay.rate}
  \Gamma(\text{H} \to b\overline{b})
  = \Gamma_0^{b}(m_b^2)\,
     \frac{\widetilde{R}_\text{\tiny S}(M_{H}^2)}
       {3\,m_b^2}\,,
\end{eqnarray}
with $\Gamma_0^{b}(m_b^2)=3\,G_F\,M_H\,m_{b}^2/{4\sqrt{2}\pi}$,
$m_b$ and $M_H$ are the pole mass of the $b$-quark and the mass of
the Higgs boson, respectively,
$\widetilde{R}_\text{\tiny S}(M_{H}^2)
 = \overline{m}^2_{b}(M_{H}^2)\,R_\text{S}(M_{H}^2)$
and
$R_\text{\tiny S}(s)$
is the $R$-ratio for the scalar correlator,
see for details in~\cite{BMS-APT,BCK05}.

In the one-loop FAPT this generates the following
non-power series expansion:\footnote{%
Appearance of denominators $\pi^n$ in association
with the coefficients $\tilde{d}_n$
is due to $d_n$ normalization.}
\begin{eqnarray}
 \label{eq:R_S-MFAPT.1L}
  \widetilde{\mathcal R}_\text{\tiny S}^{\,(1)}[L]
   =  3\,\hat{m}_{(1)}^2\,
      \Bigg\{\mathfrak A_{\nu_{0}}^{\text{\tiny glob}}[L]
          + d_1^\text{\,\tiny S}\,\sum_{n\geq1}
             \frac{\tilde{d}_{n}^\text{\,\tiny S}}{\pi^{n}}\,
              \mathfrak A_{n+\nu_{0}}^{\text{\tiny glob}}[L]
      \Bigg\}\,,
\end{eqnarray}
where $\hat{m}_{(1)}^2=8.53\pm0.09$~GeV$^2$ is the RG-invariant
of the one-loop $\overline{m}_{b}^2(\mu^2)$ evolution,
written in the following way:
$\overline{m}_{b}^2(\mu^2) = \hat{m}_{(1)}^2\,\alpha_{s}^{\nu_{0}}(\mu^2)$
with $\nu_{0}=2[\gamma_0/b_0]_{N_f=5}=1.04$,
and
$\gamma_0$ is the leading-order quark-mass anomalous dimension.
This value $\hat{m}_{(1)}^2$ has been obtained
using the one-loop relation~\cite{ChSt99,ChSt00}
between the pole mass of the $b$ quark, $m_b$,
and the value of the running mass at the scale
$\mu_{*}=\overline{m}_b(\mu_{*}^2)$,
which we call $\overline{m}_b(\overline{m}_b^2)$.
Here we also extract the value of $d_1^\text{\,\tiny S}=17\,C_\text{F}/4=17/3$
out of higher perturbative coefficients,
so that
$\tilde{d}_{n}^\text{\,\tiny S}=3\,d_{n}^\text{\,\tiny S}/17$
and
$\tilde{d}_{1}^\text{\,\tiny S}=1$.

In the two-loop case we obtain
\begin{subequations}
\begin{eqnarray}
 \label{eq:mb.Evo.2L}
  \overline{m}_b^2(\mu^2)
  &\!=\!& \hat{m}_{(2)}^2
     \alpha_{s}^{\nu_0}(\mu^2)
      \left[1 + \frac{c_1\,b_0\,\alpha_{s}(\mu^2)}{4\pi^2}\right]^{\nu_1}\,;\\
 \label{eq:R_S-MFAPT.2L}
  \widetilde{\mathcal R}_\text{\tiny S}^{\,(2)}[L]
   &\!=\!& 3\hat{m}_{(2)}^2\,
      \Bigg\{\mathfrak B_{\nu_{0},\nu_{1}}^{(2);\text{glob}}[L]
           + d_1^\text{\,\tiny S}\,\sum_{n\geq1}
              \frac{\tilde{d}_{n}^\text{\,\tiny S}}{\pi^{n}}\,
               \mathfrak B_{n+\nu_{0},\nu_{1}}^{(2);\text{glob}}[L]
      \Bigg\}\,,
\end{eqnarray}
\end{subequations}
where the RG-invariant mass $\hat{m}_{(2)}^2=8.22\pm0.09~\text{GeV}^2$,
while the value of $\nu_0$ is the same as in the one-loop case,
and $\nu_1=2\,[(\gamma_1\,b_0-\gamma_0\,b_1)/(b_0\,b_1)]_{N_f=5}=1.86$
($\gamma_1$ is the next-to-leading-order quark-mass anomalous dimension).
Note that we determined both values,
$\hat{m}_{(1)}^2$ and $\hat{m}_{(2)}^2$,
using the estimates for
$\overline{m}_b(\overline{m}_b^2)$
derived by Penin and Steinhauser~\cite{PeSt02}.

In order to estimate now
the importance of higher-order corrections,
we need to construct
some model
for the generating function $P(t)$.
We use a Lipatov-like model~\cite{BM08}
with $c=2.4$, $\beta=-0.52$,
which generates factorially growing coefficients
$\tilde{d}_{n}^\text{\,\tiny S}$:
\begin{subequations}
\label{eq:Higgs.Model}
\begin{eqnarray}
  \tilde{d}_{n}^\text{\,\tiny S}
   \!\!&\!\!=\!\!&\!\! c^{n-1}\frac{\Gamma (n+1)+\beta\,\Gamma (n)}{1+\beta}\,,
 \\
  P_\text{\tiny S}(t)
   \!\!&\!\!=\!\!&\!\! \frac{(t/c)+\beta}{c\,(1+\beta)}\,e^{-{t/c}}\,.
\end{eqnarray}
\end{subequations}
this model gives a very good prediction for
$\tilde{d}_{n}^\text{\,\tiny S}$ with $n=2, 3, 4$,
calculated in QCD PT~\cite{BCK05}:
$7.50$, $61.1$, and  $625$
in comparison with
$7.42$, $62.3$, and  $620$.
Moreover,
it predicts the value
$\tilde{d}_{5}^\text{\,\tiny S}=7826$
which is in a good agreement
with the PMS~\cite{KaSt95,ChKS97} prediction
$\tilde{d}_{5}^\text{\,\tiny PMS}=7782$,
obtained in~\cite{BMS10}.

Then, we apply the FAPT resummation technique
to estimate
how good is the $N$th order truncation
\begin{subequations}
\begin{eqnarray}
 \label{eq:FAPT.trunc}
  \Gamma_{H\to b\overline{b}}^{\,(1)}[L;N]
  = \Gamma_0^{b}\left(\hat{m}_{(1)}^2\right)\,
     \left[{\mathfrak A}_{\nu_{0}}^{\text{\tiny glob}}[L]
          + d_1^\text{\,\tiny S}\,\sum_{n=1}^{N}
             \frac{\tilde{d}_{n}^\text{\,\tiny S}}{\pi^{n}}\,
              {\mathfrak A}_{n+\nu_{0}}^{\text{\tiny glob}}[L]
   \right]
\end{eqnarray}
in approximating the whole one-loop width
$\Gamma_{H\to b\overline{b}}^{(1)}[L]$
in the range $L\in[12.4,13.5]$
which corresponds to the range
$M_H\in[100,172]$~GeV
with $\Lambda^{N_f=3}_{\text{QCD}}=201$~MeV
and ${\mathfrak A}^{\text{\tiny glob}}_{1}(m_Z^2)=0.1226$.
In this range,
we have $L_5<L<L_6=\ln(m_t^2/\Lambda_3^2)$
and hence
\begin{eqnarray}
 \Gamma_{H\to b\overline{b}}^{(1);\infty}[L]
  = \Gamma_0^{b}\left(\hat{m}_{(1)}^2\right)\,
    \left\{\mathfrak A^{\text{\tiny glob}}_{\nu_0}[L]
     + \frac{d_{1}^\text{\,\tiny S}}{\pi}\,
        \left\langle\!\!\!\left\langle{
         \overline{\mathfrak A}_{1+\nu_{0}}
          \Big[L\!+\!\lambda_5\!-\!\frac{t}{\pi\beta_5};5\Big]
       + \Delta_{6}\overline{\mathfrak A}_{1+\nu_{0}}
          \left[\frac{t}{\pi}\right]
        }\right\rangle\!\!\!\right\rangle_{\!\!P_{\nu_{0}}^\text{\,\tiny S}}
    \right\}~
 \label{eq:R_S.Sum}
\end{eqnarray}
\end{subequations}
with $P_{\nu_{0}}^\text{\,\tiny S}(t)$ defined via Eqs.\ (\ref{eq:P_nu(t)})
and (\ref{eq:Higgs.Model}).
We analyze the accuracy of the truncated FAPT expressions (\ref{eq:FAPT.trunc})
and compare them with the total sum (\ref{eq:R_S.Sum})
using relative errors
$\Delta_N[L]=1-\Gamma_{H\to b\overline{b}}^{\,(1)}[L;N]/\Gamma_{H\to b\overline{b}}^{(1);\infty}[L]$.
In Fig.~1
we show these errors for $N=2$, $N=3$, and $N=4$
in the analyzed range of $L\in[11,13.7]$.
We see that already $\Gamma_{H\to b\overline{b}}^{\,(1)}[L;2]$
gives an accuracy of the order of 2.5\%,
whereas $\Gamma_{H\to b\overline{b}}^{\,(1)}[L;3]$
is of the order of 1\%.
\begin{figure}[h!]
 \centerline{\includegraphics[width=0.485\textwidth]{
  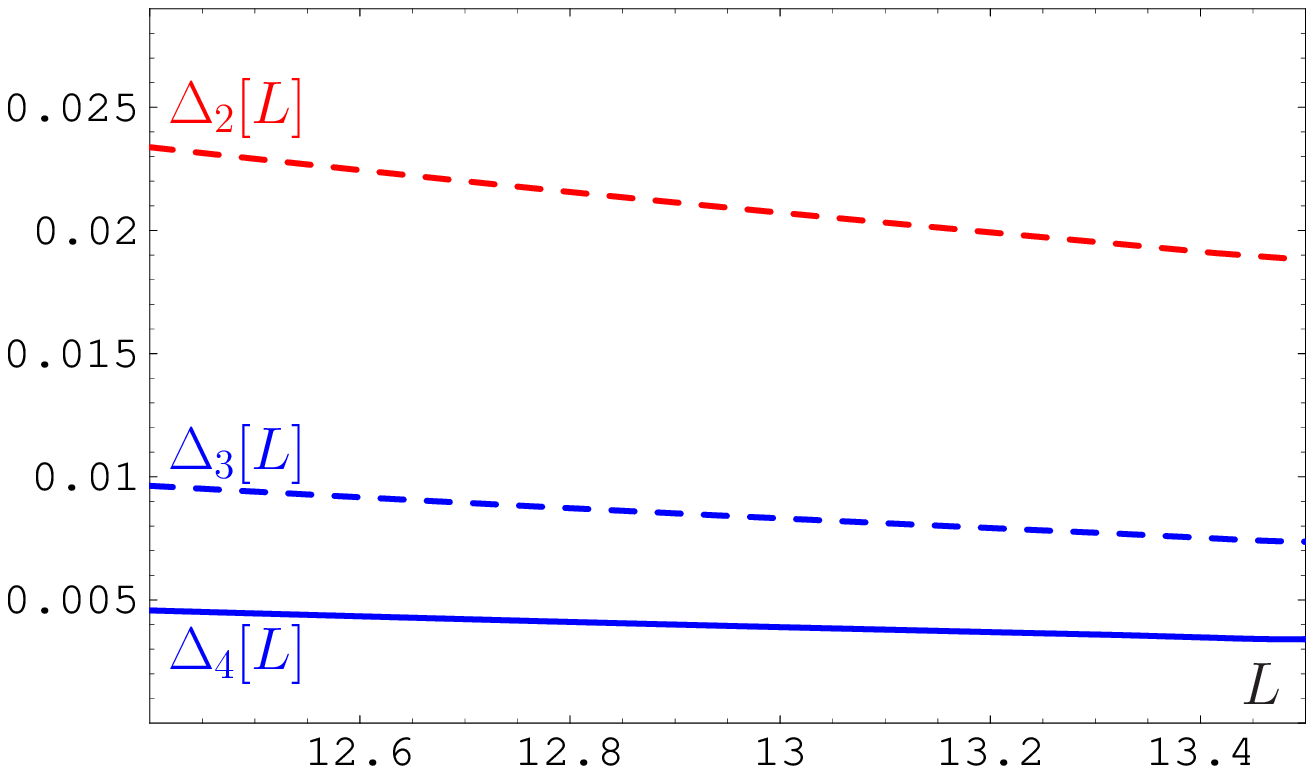}~~~\includegraphics[width=0.475\textwidth]{
  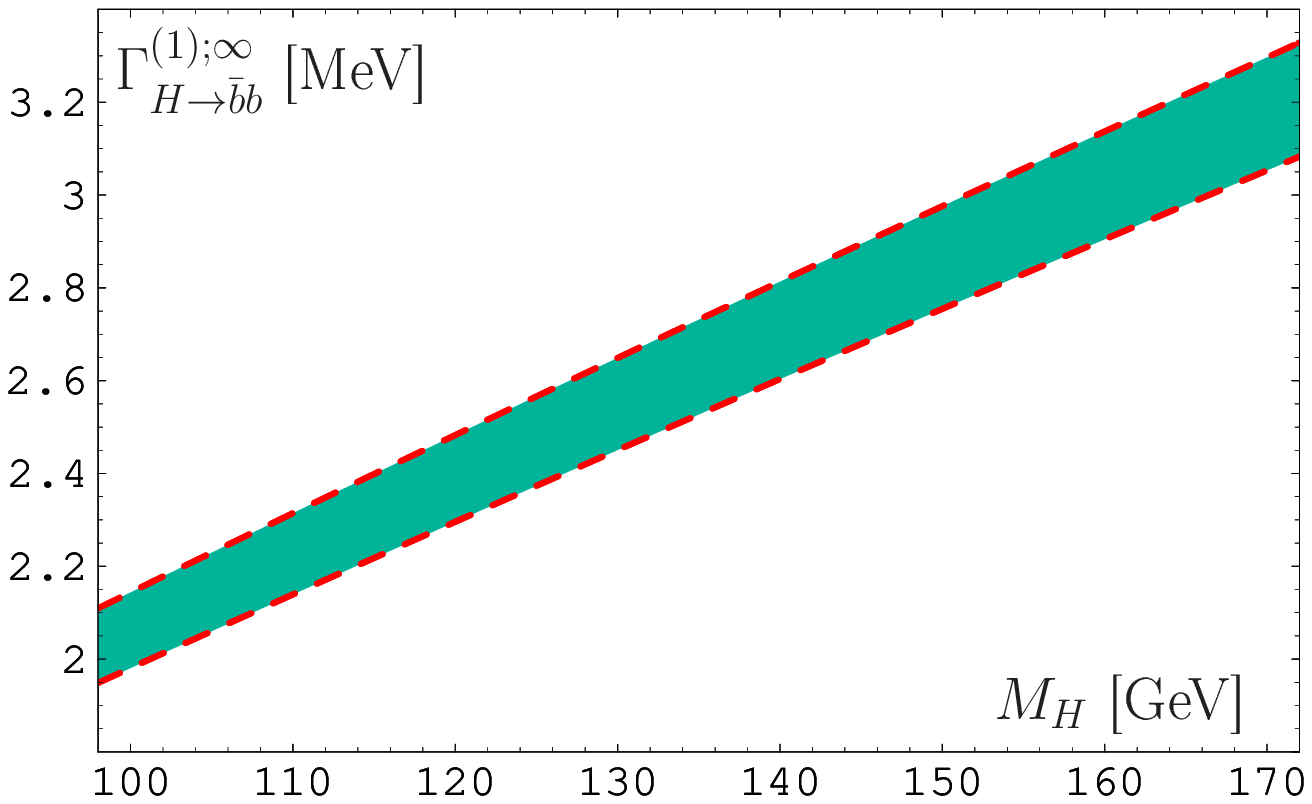}}
  \caption{Left panel: The relative errors $\Delta_N[L]$, $N=2, 3$
   and $4$, of the truncated FAPT in comparison with
   the exact summation result, Eq.\ (\ref{eq:R_S.Sum}).
   Right panel: The width $\Gamma_{H\to b\overline{b}}$ as a function
   of the  Higgs boson mass $M_{H}$ in the resummed FAPT.
   The width of the shaded strip is due to the overall uncertainties,
   induced by the uncertainties of the resummation procedure
   and the pole mass error-bars.
   Both panels show the results obtained in the one-loop FAPT~\cite{BMS10}.
   \label{fig:resum.1L}}
\end{figure}
Looking in the left panel of Fig.\,1,
we understand that in order to have an accuracy better
than 0.5\%,
we need to take into account the 4-th correction.
We also verified that the uncertainty due to $P(t)$-modelling
is small, $\lesssim0.6\%$,
while the $\overline{m}_b(\overline{m}_b^2)$-induced
uncertainty is of the order of $2\%$.
The overall uncertainty then is of the order of $3\%$,
see in the right panel of Fig.\,1.

\begin{figure}[t!]
 \centerline{\includegraphics[width=0.49\textwidth]{
  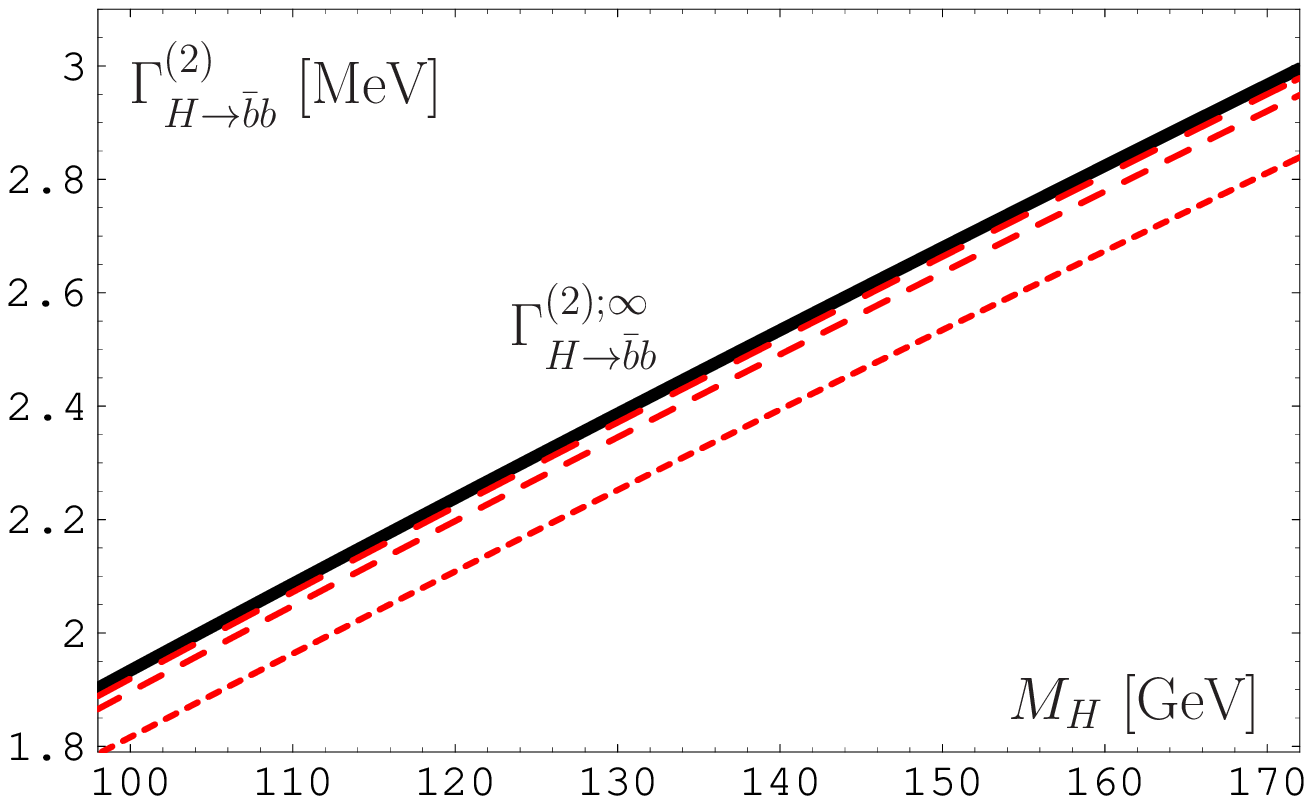}~\includegraphics[width=0.49\textwidth]{
  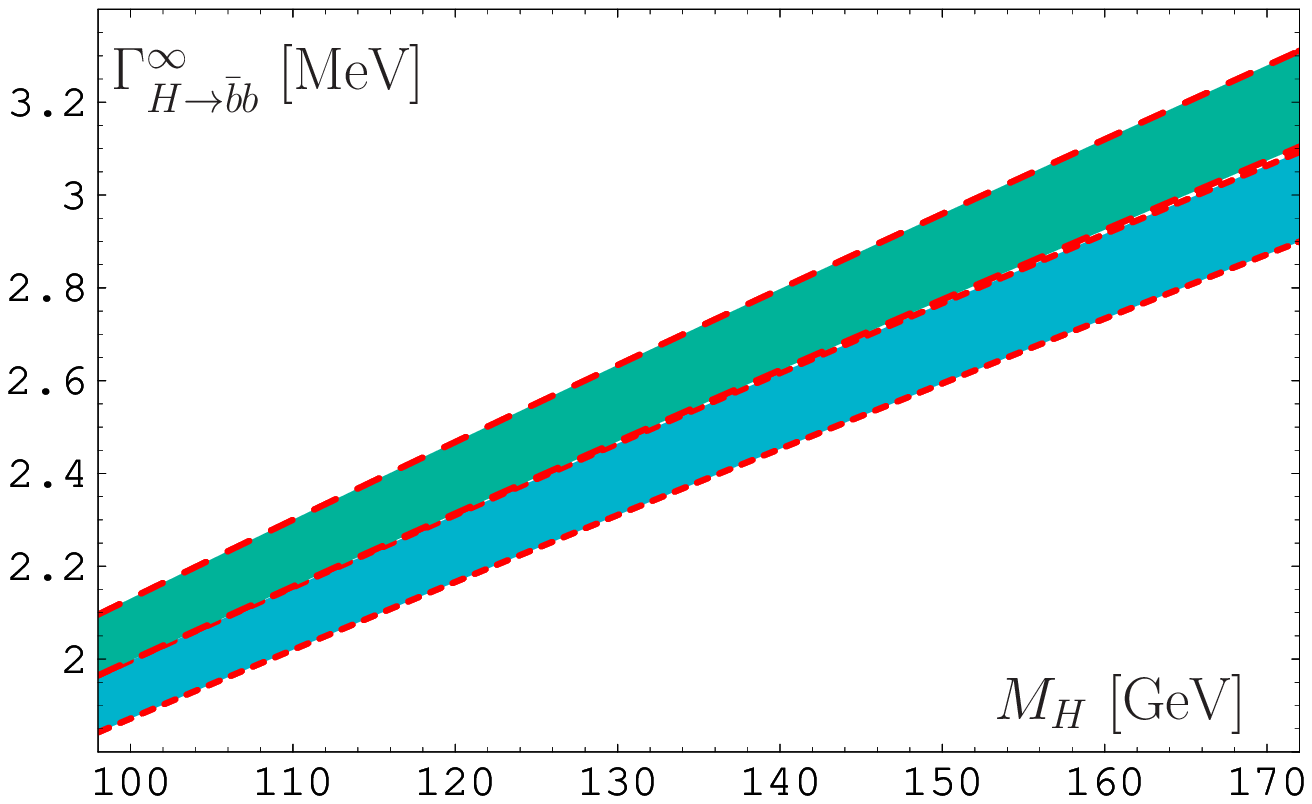}\vspace*{-1mm}}
  \caption{Left panel:~The two-loop width $\Gamma_{H\to b\overline{b}}$ as a
    function of the Higgs-boson mass $M_{H}$ in the resummed (black
    solid line) and the truncated (at the order $N$) FAPT~\cite{BMS10}.
    Here, the short-dashed line corresponds to $N=1$, the dashed one to
    $N=2$, and the long-dashed line to $N=3$.
    Right panel:~The two-loop width $\Gamma_{H \to b\overline{b}}^{\infty}$
    as a function of the Higgs-boson mass $M_H$ in the resummed FAPT
    is shown (lower strip).
    The upper strip shows the corresponding one-loop result.
    \label{fig:resum.2L}}
\end{figure}

Qualitatively,
the same picture is reproduced
at the two-loop order~\cite{BMS10}.
In the left panel of Fig.\ \ref{fig:resum.2L},
we discuss the convergence properties of the decay widths,
truncated at the order $N$,
relative to the resummed two-loop result
$\Gamma_{H \to b\overline{b}}^{(2);\infty}(M_H)$.
We see that our conclusions drawn
from the one-loop analysis remain valid.
Indeed, $\Gamma_{H\to b\overline{b}}^{(2)}(M_H;2)$
deviates from the resummed quantity by not more than 2\%,
whereas $\Gamma_{H\to b\overline{b}}^{(2)}(M_H;3)$
reaches an even higher precision level of the order of 0.7\%.

In the right panel of Fig.\ \ref{fig:resum.2L},
we show the results for the decay width
$\Gamma_{H \to b\overline{b}}^{\infty}(M_H)$
in the resummed two-loop FAPT,
varying the mass in the interval
$\hat{m}_{(2)}=8.22\pm0.13$~GeV
according to the Penin--Steinhauser estimate
$\overline{m}_b(\overline{m}_b^2)=4.35\pm0.07$~GeV~\cite{PeSt02}.
Comparing this outcome with the one-loop result---upper strip in the
same panel of this figure---reveals a 5\% reduction of the two-loop
estimate.
This reduction consists of two parts:
one part ($\approx+7\%$) is due to the difference $\hat{m}_{(1)}^2-\hat{m}_{(2)}^2$
in both approximations,
while the other ($\approx-2\%$) comes from the
difference in the values of $R_{S}(M_H)$ in the one- and the
two-loop approximations.

\section{Application to the vector correlator Adler function}
\label{sec:App.Adler}

  Now we analyze the Adler function of the vector correlator.
We model the generating function of the perturbative coefficients
$d_n$~\cite{BCK0810}
(note that here $d_1=1$, so that $\tilde{d}_{n}=d_n$)
\begin{eqnarray}
  D_\text{V}[L]
  = d_0
  + \sum_{n=1}^{\infty}d_n\,
  \left(\frac{\alpha_s[L]}{\pi}
  \right)^{n}
 \label{eq:D_V}
\end{eqnarray}
using a generalized Lipatov-like model:
\begin{subequations}
\label{eq:Vector.Model}
\begin{eqnarray}
\label{eq:Vector.d_n.Model}
 d_n^\text{V}
  &=& c^{n-1}\,
    \frac{\delta^{n+1}-n}
         {\delta^2-1}\,\Gamma(n)\,,\\
\label{eq:Vector.P(t).Model}
 P_\text{V}(t)
  &=&
   \frac{\delta\,e^{-t/c\delta} - (t/c)\,e^{-t/c}}
        {c\left(\delta ^2-1\right)}\, .
\end{eqnarray}
\end{subequations}
Our predictions, obtained with this generating function
by fitting the two known coefficients $\tilde{d}_2$ and $\tilde{d}_3$,
have been included in Table \ref{Tab:d_n.Adler}
(see~\cite{BMS10}).
We observe a good agreement between our estimate
$d_4^\text{V}=27.1$
and the value 27.4, calculated recently by Chetyrkin \textit{et al.}
in Ref.\ \cite{BCK0810}.
Would we use instead a fitting procedure,
which would take into account the fourth-order coefficient $d_4$
in order to predict $d_5$,
we should readjust the model parameters
in (\ref{eq:Vector.Model})
to the new values
$\left\{c=3.5526,~\delta=1.32453\right\}$.
In order to explore to what extent the exact knowledge
of the
higher-order coefficients $d_n$ is important,
we employed our model (\ref{eq:Vector.Model}) in~\cite{BMS10}
with different values of the parameters:
$c=3.63$ and $\delta=1.3231$.
These values were selected in order to reproduce
the exact value of the coefficient
$d_4=27.4$ by the value,
close to the Naive Non-Abelization  (NNA)
prediction~\cite{BroadKa}.
The difference between the resummed results of the two models
in the region of $N_f=4$
appears to be very small --- of the order of $0.2\%$.
\begin{table}[ht]
\caption{Coefficients $d_n$ for the Adler-function series with $N_f=4$.
The numbers in the square brackets denote the lower and the upper limits
of the Improved NNA (INNA) estimates.\label{Tab:d_n.Adler}}
\centerline{
\begin{tabular}{|c|ccccc|}\hline \hline
 PT coefficients     &~$d_1\vphantom{^{|}_{|}}$~
                             &~$d_2$~
                                    &~$d_3$~
                                           &~$d_4$~
                                                  &~$d_5$~
\\ \hline \hline
pQCD results with $N_f=4$ \cite{BCK0810}
                       & $1\vphantom{^{|}_{|}}$
                             &~1.52~&~2.59~&~27.4~&~---~
\\ \hline \hline
Model (\ref{eq:Vector.Model}) with $c=3.555,~\delta=1.3245$~
                       & $1\vphantom{^{|}_{|}}$
                             &~1.52~&~2.59~&~27.1~&~2024~
\\ \hline
Model (\ref{eq:Vector.Model}) with $c=3.553,~\delta=1.3245$
                       & $1\vphantom{^{|}_{|}}$
                             &~1.52~&~2.60~&~27.3~&~2025~
\\ \hline
Model (\ref{eq:Vector.Model}) with $c=3.630,~\delta=1.3231$
                       & $1\vphantom{^{|}_{|}}$
                             &~1.53~&~2.26~&~20.7~&~2020~
\\ \hline
``INNA'' prediction of~~\cite{BMS10}
                       & $1\vphantom{^{|}_{|}}$
                             &~1.44~&~$[3.5,9.6]$~
                                            &~$[20.4,48.1]$~
                                                   &~$[674,2786]$~
\\ \hline \hline
\end{tabular}}
\end{table}
We also show in Table~\ref{Tab:d_n.Adler}
the $d_n$ estimates obtained
by using the Improved NNA (INNA) approximation,
developed in our last paper~\cite{BMS10}
by taking into account not only
the $b_0^n\alpha_s^n$ terms,
but also the $b_1^kb_0^{n-2k}\alpha_s^{n}$ ones.
This technique produces predictions for higher-order
coefficients in terms of intervals for the $d_n $values.
We see that our model predictions for the $d_5$ coefficient
are all inside the INNA interval.

Then, we have
\begin{subequations}
\begin{eqnarray}
 \label{eq:R_V.APT.exact}
  D_\text{V}^{\text{APT}}(Q^2)
  &=& 1 + \sum_{n\geq1}
           \frac{d^\text{V}_{n}}{\pi^{n}}\,
            \mathcal A_{n}^{\text{\tiny glob}}(Q^2)\,,
\\
 \label{eq:R_V.APT.trunc}
  D_\text{V}^{\text{APT}}(Q^2;N)
  &=& 1 + \sum_{n=1}^{N}
           \frac{d^\text{V}_{n}}{\pi^{n}}\,
            \mathcal A_{n}^{\text{\tiny glob}}(Q^2)\,.
\end{eqnarray}
\end{subequations}
\begin{figure}[b!]
 \centerline{\includegraphics[width=0.5\textwidth]{
  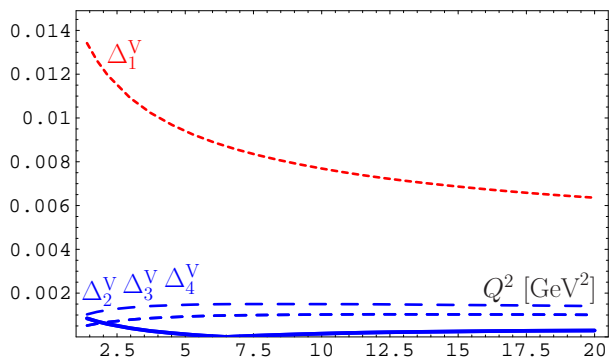}\vspace*{-1mm}}
   \caption{The relative errors $\Delta^\text{V}_N(Q^2)$ evaluated for
    different values of $N$:
    $N=1$ (short-dashed red line),
    $N=2$ (solid blue line),
    $N=3$ (dashed blue line), and
    $N=4$ (long-dashed blue line) of the truncated APT given by
    Eq.\ (\ref{eq:R_V.APT.trunc}), in comparison with the exact
    result of the APT resummation procedure.
   \label{fig:Adler}}
\end{figure}
The global-APT resummation result for
$D_\text{V}^{\text{APT}}(Q^2)$
can be estimated using our resummation formulas from~\cite{AB08,BM08,BMS10},
which we shortly explained in Sect.~\ref{sec:Resum.FAPT}.
We show in Fig.\ \ref{fig:Adler} the relative errors
$\Delta_N^\text{V}(Q^2)=1
 - D_\text{V}^{\text{APT}}(Q^2;N)/D_\text{V}^{\text{APT}}(Q^2)$
evaluated in the range
$Q^2\in[2,20]$~GeV$^{2}$ for four values $N=1, 2, 3, 4$.
We observe from this figure that already
$D_\text{V}^{\text{APT}}(Q^2;1)$
provides an accuracy in the vicinity of 1\%,
whereas $D_\text{V}^{\text{\tiny APT}}(Q^2;2)$
is smaller then 0.1\%
in the interval $Q^2=1-20$~GeV$^2$.
This means that there is no real need to calculate further corrections:
at the level of an accuracy of  the order of 1\%
it is quite enough to take into account
only the terms up to $d_2$.
Again, we verified that this conclusion is stable
with respect to the variation of the model $P_\text{V}(t)$ parameters.
This conclusion is in some sense surprising:
The best order of truncation of the FAPT series
in the region $Q^2=2-20$~GeV$^2$
is reached by the N$^2$LO approximation,
i.e., by keeping just the $d_2$-term.

\section{Conclusions}
\label{sec:Concl}
We described here the resummation approach
in the global versions of both APT and FAPT
for the one- and two-loop running of the effective QCD coupling.
We argued that this approach produces finite answers,
provided the generating function $P(t)$
of the perturbative coefficients $d_n$ is known.
We considered two applications of our approach
to the physical problems of current interest.

In the case of the Higgs boson decay, $H\to\bar{b}b$,
the main conclusion is that
to achieve a truncation accuracy of the order of 1\%
for the width  $\Gamma_{H\to b\overline{b}}(M_H)$
it is enough to take into account
up to the third correction ---
in complete agreement with Kataev\&Kim~\cite{KK08}.
In the case of the vector-current Adler function,
$\mathcal D_\text{V}(Q^2)$,
the truncation accuracy is of the order 0.1\%,
already reached  at N$^2$LO,
i.\,e.,
by taking into account up to the second-order correction.

In our approach the knowledge of higher-order coefficients, $d_4$,
is needed in order to construct a reliable model
for the generating functions $P(t)$.
Meanwhile, we showed
that the exact values of still higher coefficients,
$d_5$, are not so important.

\section*{Acknowledgements}
I wish to thank Sergey Mikhailov and Nico Stefanis
for a fruitful collaboration,
and Konstantin Chetyrkin, Andrey Grozin, Andrey Kataev and Alexey Pivovarov
for useful remarks and discussions.
This work was supported in part by
the Russian Foundation for Fundamental Research,
grant No.\ ü~08-01-00686,
the BRFBR--JINR Cooperation Programme,
contract No.\ F10D-001, and
the Heisenberg--Landau Programme under Grant 2010.


\begin{thebibliography}{1}

\bibitem{BLS60}
 N.~N. Bogolyubov, A.~A. Logunov, and D.~V. Shirkov,
  Soviet Physics JETP \textbf{10} (1960)  574.

\bibitem{Rad82}
 A.~V. Radyushkin,
  JINR Rapid Commun. \textbf{78} (1996) 96;
  [JINR Preprint, E2-82-159, 26 Febr. 1982;
    arXiv: hep-ph/9907228].

\bibitem{KP82}
 N.~V. Krasnikov and A.~A. Pivovarov,
  Phys. Lett. \textbf{B116} (1982) 168.

\bibitem{DMW96}
 Y.~L. Dokshitzer, G. Marchesini, and B.~R. Webber,
  Nucl. Phys. \textbf{B469},  93  (1996).

\bibitem{JS95-349}
 H.~F. Jones and I.~L. Solovtsov,
  Phys. Lett. \textbf{B349},  519  (1995).

\bibitem{JS95-357}
 H.~F. Jones, I.~L. Solovtsov, and O.~P.~Solovtsova,
  Phys. Lett. \textbf{B357},  441  (1995).

\bibitem{BB95}
 M. Beneke and V.~M. Braun,
  Phys. Lett. \textbf{B348} (1995)  513.

\bibitem{BBB95}
 P. Ball, M. Beneke, and V.~M. Braun,
  Nucl. Phys. \textbf{B452} (1995) 563.

\bibitem{SS}
 D.~V. Shirkov and I.~L. Solovtsov,
  JINR Rapid Commun. \textbf{2} [76] (1996) 5;
  Phys. Rev. Lett. \textbf{79} (1997) 1209;
  Theor. Math. Phys. \textbf{150} (2007) 132.

\bibitem{Sim01}
 Y.~A. Simonov,
  Phys. Atom. Nucl. \textbf{65} (2002) 135.

\bibitem{DVS-Global}
 D.~V. Shirkov,
  Theor. Math. Phys. \textbf{127},  409  (2001);
  Eur. Phys. J. \textbf{C22},  331  (2001).

\bibitem{KS01}
 A.~I. Karanikas and N.~G. Stefanis,
  Phys. Lett. \textbf{B504} (2001) 225;
              \textbf{B636}, 330 (2006).

\bibitem{SSK9900}
 N.~G. Stefanis, W. Schroers, and H.-C. Kim,
  Phys. Lett. \textbf{B449}  (1999)  299;
  Eur. Phys. J. \textbf{C18} (2000) 137.

\bibitem{Ste02}
 N.~G. Stefanis,
  Lect. Notes Phys. \textbf{616} (2003) 153.

\bibitem{BPSS04}
 A.~P. Bakulev, K. Passek-Kumeri\v{c}ki, W. Schroers, and N.~G. Stefanis,
  Phys. Rev. \textbf{D70}  (2004)  033014.

\bibitem{BMS-APT}
 A.~P. Bakulev, S.~V. Mikhailov, and N.~G. Stefanis,
  Phys. Rev. D\textbf{72} (2005) 074014, 119908(E);
  \textit{ibid.} \textbf{75} (2007) 056005;
              \textbf{77} (2008) 079901(E).

\bibitem{BKS05}
 A.~P. Bakulev, A.~I. Karanikas, and N.~G. Stefanis,
  Phys. Rev. D\textbf{72} (2005) 074015.

\bibitem{AB08}
 A.~P. Bakulev
  Phys. Elem. Part. Nucl. \textbf{40} (2009) 715.

\bibitem{Ste09}
 N.~G. Stefanis, arXiv:0902.4805 [hep-ph].

\bibitem{MS04}
  S.~V. Mikhailov,
  JHEP \textbf{06} (2007) 009 [hep-ph/0411397].

\bibitem{Neu95prd}
 M. Neubert,
  Phys. Rev. \textbf{D51} (1995) 5924.

\bibitem{CV06}
 G. Cvetic and C. Valenzuela,
  Phys. Rev. \textbf{D74} (2006) 114030 [hep-ph/0608256].

\bibitem{BM08}
 A.~P. Bakulev and S.~V. Mikhailov,
  in \textit{Proc. Int. Seminar on Contemp.
  Probl. of Part. Phys.,
  dedicated to the memory of I.~L.~Solovtsov,
  Dubna, Jan. 17--18, 2008.},
  Eds. A.~P.~Bakulev \textit{et~al.}
  (JINR, Dubna, 2008),
   pp.\ 119--133
  (arXiv:0803.3013).

\bibitem{BMS10}
 A.~P. Bakulev, S.~V. Mikhailov, and N.~G. Stefanis,
  JHEP \textbf{1006} (2010) 085.

\bibitem{BCK05}
 P.~A. Baikov, K.~G. Chetyrkin, and J.~H. K{\"u}hn,
  Phys. Rev. Lett. \textbf{96} (2006) 012003.

\bibitem{ChSt99}
 K.~G. Chetyrkin and M. Steinhauser,
  Phys. Rev. Lett. \textbf{83},  4001  (1999).

\bibitem{ChSt00}
 K.~G. Chetyrkin and M. Steinhauser,
  Nucl. Phys. \textbf{B573} (2000) 617.

\bibitem{PeSt02}
 A.~A. Penin and M. Steinhauser,
  Phys. Lett. \textbf{B538} (2002) 335.

\bibitem{KaSt95}
 A.~L. Kataev and V.~V. Starshenko,
  Mod. Phys. Lett. \textbf{A10} (1995) 235.

\bibitem{ChKS97}
 K.~G. Chetyrkin, B.~A. Kniehl, and A. Sirlin,
  Phys. Lett. \textbf{B402} (1997) 359.

\bibitem{BCK0810}
 P.~A. Baikov, K.~G. Chetyrkin, and J.~H. K{\"u}hn,
  Phys. Rev. Lett. \textbf{101} (2008) 012002;
  \textit{ibid.} \textbf{104} (2010) 132004.

\bibitem{BroadKa}
 D.~J. Broadhurst and A.~L. Kataev,
  Phys. Lett. \textbf{B315} (1993) 179;
  \textit{ibid.} \textbf{B544} (2002) 154.

\bibitem{KK08}
 A.~L. Kataev and V.~T. Kim,
  in \textit{Proc. Int. Seminar on Contemp.
  Probl. of Part. Phys.,
  dedicated to the memory of I.~L.~Solovtsov,
  Dubna, Jan. 17--18, 2008.},
  Eds. A.~P.~Bakulev \textit{et~al.}
  (JINR, Dubna, 2008),
   pp.\ 167--182
   (arXiv:0804.3992);\\
  PoS \textbf{ACAT08} (2009) 004.

\end{thebibliography}


\end{document}